\documentclass[twoside]{IEEEtran}

\makeatletter
\def\ps@headings{%
\def\@oddhead{\mbox{}\scriptsize\rightmark \hfil \thepage}%
\def\@evenhead{\scriptsize\thepage \hfil \leftmark\mbox{}}%
\def\@oddfoot{}%
\def\@evenfoot{}}
\makeatother
\usepackage{array, latexsym, graphicx,times,amsfonts}
\usepackage{url}
\usepackage{epsfig}
\usepackage{amsmath}
\usepackage{verbatim}
\usepackage{blkarray}
\usepackage{setspace}
\usepackage{subfigure}
\usepackage{multirow}

\newtheorem{theorem}{Theorem}[section]

\newtheorem{corollary}[theorem]{Corollary}

\newtheorem{ex}[theorem]{Example}
\newtheorem{definition}[theorem]{Definition}
\newtheorem{fact}[theorem]{Fact}

\begin{document}

\title{LT Codes For Efficient and Reliable Distributed Storage Systems Revisited}

\author{
\IEEEauthorblockN{Yongge Wang}
\IEEEauthorblockA{
\\
Department of SIS, UNC Charlotte\\
9201 University City Blvd., Charlotte, NC 28223, USA\\
Email: yongge.wang@uncc.edu}
}

\maketitle

\begin{abstract}
LT codes and digital fountain techniques have received significant 
attention from both academics and industry in the past few years. 
There have also been extensive interests in applying LT code techniques
to distributed storage systems such as cloud data storage in recent years. 
However, Plank and Thomason's experimental results show that
LDPC code performs well only asymptotically when the number of 
data fragments increases and it has the worst performance for small 
number of data fragments (e.g., less than $100$). In their INFOCOM
2012 paper, Cao, Yu, Yang, Lou, and Hou proposed to use 
exhaustive search approach to find a deterministic LT code
that could be used to decode the original data content correctly
in distributed storage systems.
However, by Plank and Thomason's experimental results,
it is not clear whether the exhaustive search approach will 
work efficiently or even correctly.
This paper carries out the theoretical analysis on the feasibility
and performance issues for applying LT codes to distributed storage
systems. By employing the underlying ideas of efficient 
Belief Propagation (BP) decoding process in LT codes, 
this paper introduces two classes
of codes called flat BP-XOR codes and array BP-XOR codes 
(which can be considered as
a deterministic version of LT codes). 
We will show the equivalence between the edge-colored graph model 
and degree-one-and-two encoding symbols based
array BP-XOR codes. Using this equivalence result, we are able to design
general array BP-XOR codes using graph based results. Similarly, 
based on this equivalence result, we are able to get new results 
for edge-colored graph models using results from array BP-XOR codes.
\end{abstract}

\section{Introduction}
Due to the advancement of cloud computing technologies,
there has been an increased interest for individuals and business
entities to move their data from traditional private data center
to cloud servers. Indeed, even popular storage service providers 
such as Dropbox use third party cloud storage providers such as 
Amazon's Simple Storage Service (S3) for data storage.

With the wide adoption of cloud computing and storage technologies,
it is important to consider data security and reliability issues 
that are strongly related to the underlying storage services.
Though it is interesting to consider general data security issues
in cloud computing environments, this paper will concentrate on
the basic question of reliable data storage in the cloud
and specifically the coding techniques for data storage in the cloud. 
There has been extensive research in reliable data storage 
on disk drives. For example, redundant array of independent disks
(RAID) techniques have been proposed and widely adopted 
to combine multiple disk drive components into a logical unit
for better resilience, performance, and capacity. 
The well known solutions to address the data storage 
reliability are to add data redundancy to multiple drivers.
There are basically two ways to add the redundancy:
data mirror (e.g., RAID 1) and data stripping with 
erasure codes (e.g., RAID 2 to RAID 6).
Though data mirror (or data replication) provides 
the straightforward way for simple data management
and repair of data on corrupted drives, it is very expensive to
implement and deploy due to its high demand for redundancy. 
In addition to data replication techniques,
erasure codes can be used to achieve the required data reliability 
level with much less data redundancy. Note that though error 
correcting codes (e.g., Reed-Solomon codes)  could also be used 
for reliable data storage and correcting errors from failed disk drives, 
it is normally not used for data storage since it needs expensive 
computation for both encoding and decoding processes.

Erasure codes that have been used for reliable data storage systems
are mainly binary linear codes which are essentially XOR-operation 
based codes.  For example, flat XOR codes are erasure codes 
in which parity disks are calculated as the XOR of 
some subset of data disks.  Though it is desirable 
to have MDS (maximal distance separable) flat XOR codes,
it may not be available for all scenarios. Non-MDS codes have also
been used in storage systems (e.g., the replicated RAID 
configurations such as RAID 10, RAID 50, and RAID 60). 
However, we have not seen any systematic research in 
designing non-MDS codes with flat XOR operations for storage systems.

In order to achieve better fault tolerance with minimal redundancy
in data storage systems, there has also been active research 
in XOR based codes 
which are not necessarily  flat XOR codes. For example,
Blaum, Brady, Bruck, and Menon \cite{DBLP:journals/tc/BlaumBBM95}  
proposed the array code EVENODD for tolerating two disk faults 
and correcting one disk errors. 
Blaum,  Bruck, and  Vardy \cite{Blaum96mdsarray} 
and Huang \cite{Huang05anefficient} have extended the construction
of EVENODD code to general codes for tolerating three disk faults.
Other non-flat XOR based codes include
(but are not limited to) $[2k,k,d]$ chain code,
Simple Product Code (SPC \cite{Elias}), Row-Diagonal Parity 
(RDP \cite{DBLP:conf/fast/CorbettEGGKLS04}), and X-code
\cite{Xu99x-code:mds}.

The techniques that we have discussed above have been originally designed
for data storage on disk drives and in storage area networks. It may not be
directly applicable to distributed storage services such as the storage cloud.
There have been many researches addressing the data storage reliability
issues in distributed environments. For example,
Weatherspoon and Kubiatowicz \cite{Weatherspoon:2002:ECV:646334.687814}
have compared erasure coding based solutions and replication based solutions 
for reliable distributed data storage systems.

Based on the seminal work of low-density parity-check (LDPC) codes
by Gallager \cite{gallagerldpc}, several important techniques
(see, e.g., Luby, Mitzenmacher, Shokrollahi, Spielman, and Stemann
\cite{Luby:1997:PLC:258533.258573})  have 
been developed for networked and communication systems such as
Digital Fountain for content streaming.  There have also been extensive
interests in applying digital fountain techniques (such as LT codes)
to distributed storage systems (see, e.g., Plank and Thomason \cite{ldpc}
and Cao, Yu, Yang, Lou, and Hou \cite{CaoYuYangLouHouLTcodes}).
However, it is not clear whether these applications of LT codes to
distributed storage systems have advantages over other
techniques that have been extensively adopted in disk drives 
and storage area networks such as flat XOR codes or array XOR codes.

Plank and Thomason's experimental results \cite{ldpc} 
show that LDPC code performs well only asymptotically when the number of 
data packages increases and it has the worst performance for small 
number of data fragments (e.g., less than $100$). 
Cao, Yu, Yang, Lou, and Hou \cite{CaoYuYangLouHouLTcodes}
proposed to use exhaustive search approach
to find a deterministic LT code
that could be used to decode the original data content correctly
in distributed storage systems.
However, by Plank and Thomason's experimental results,
it is not clear whether the exhaustive search approach will 
work efficiently or even correctly.
In this paper, we carry out a theoretical analysis on the feasibility
and performance issues for applying LT codes to distributed storage
systems. By employing the underlying ideas of efficient 
Belief Propagation (BP) decoding process in LT codes
(\cite{DBLP:conf/focs/Luby02}), 
we introduce two classes of codes called flat BP-XOR codes and 
array BP-XOR codes. The BP-XOR codes and array BP-XOR codes can be considered as
a deterministic version of LT codes
though flat BP-XOR codes are different from LT codes.
Edge-colored graph models were introduced by
Wang and Desmedt in \cite{Wang:2011:EGA:1975023.1975459}
to model homogeneous faults in networks.
We will show the equivalence between the edge-colored graph model 
and degree-one-and-two encoding symbol based
array BP-XOR codes. Using this equivalence result, we are able to design
general array BP-XOR codes using graph based results. Similarly, 
based on this equivalence result, we are able to get new results 
for edge-colored graph models using results from array BP-XOR codes.
We have implemented an online 
software package for users to generate array BP-XOR code
with their own specification and to verify the validity of 
their array BP-XOR codes (see \cite{bpxorurl}).

The structure of this paper is as follows. In Section \ref{reviewtech},
we briefly review several coding techniques proposed 
for distributed storage systems. Sections \ref{challengesec}  presents
the challenges in applying LT codes to distributed storage systems. 
Section \ref{ltcodesec} introduces flat BP-XOR codes
for distributed storage systems and investigates the necessary
and sufficient bounds for the existence of the codes.
Section \ref{arraybpxorsec} introduces array BP-XOR codes 
for distributed storage systems and establish the equivalence between
edge-colored graph models and array BP-XOR codes with 
degree one and two encoding symbols. 
Section \ref{arraybpxorexamplesec} presents constructions
of array BP-XOR codes from graph based results (e.g., perfect
one factorization of complete graphs),
and Section \ref{comparisonsec} proves a theorem 
on the limitation of array BP-XOR codes using only 
degree one and two encoding symbols.

\section{Coding Techniques for Distributed Storage Systems}
\label{reviewtech}
In the seminal paper \cite{Rabin:1989:EDI:62044.62050}, Rabin 
proposed the Information Dispersal Algorithm (IDA) to code
a data file into $n$ pieces that will be stored among $n$ servers
such that the recovery of the information is possible when there 
are at most $t=n-k$ failed servers (inactive but not malicious Byzantine style
servers). Rabin's scheme is essentially a kind of Reed-Solomon
codes and needs relatively expensive finite field operations for 
encoding and decoding.

Krawczyk \cite{Krawczyk:1993:DFS:164051.164075} extended
Rabin's IDA scheme to address the Byzantine style malicious servers
which may intentionally modify their pieces of the information.
Krawczyk called his scheme as Secure Information Dispersal Algorithm 
(SIDA).

There have been extensive interests in applying random linear coding
techniques to distributed storage systems. For example,
Dimakis, Ramchandran, Wu, and Suh \cite{DBLP:journals/corr/abs-1004-4438}
and Dimakis, Godfrey, Wu,  Wainwright, and Ramchandran
\cite{Dimakis:2010:NCD:1861840.1861868}
used information flow graphs
and random linear coding to achieve information theoretic
minimum functional repair bandwidth 
$\gamma_{\mbox{min}}=\frac{2|F|d}{2kd-k^2+k}$.
In another word, if we divide the file $F$ into $k$ pieces, store the encoded
fragments in $n$ storage servers, and if one of these server fails, then
a new comer storage server could functionally repair the system
only if it could communicate $\gamma_{\mbox{min}}/d$ bits from
each of the $d$ surviving storage servers.

Inspired by network coding, Acedanski, Medard, and Koetter 
\cite{Acedanski05howgood} proposed to use random linear coding 
for distributed networked storage with one centralized server and 
multiple storage servers. 
Dimakis, Prabhakaran, and Ramchandran \cite{Dimakis06decentralizederasure} 
considered the problem from a different approach: There are $n$ 
storage servers and many distributed data sources (that is, data 
are not from a central location and there is no centralized server). 
Each data source node picks one out of the $n$ storage servers randomly, 
pre-routes its packet and repeat $d(k)=c\ln k$ times. Each storage server 
multiplies what it receives with coefficients selected uniformly and 
independently from $F_q$ and stores the results together with the 
coefficients. 

\section{Challenges in Applying LT codes to distributed storage systems}
\label{challengesec}
Luby \cite{DBLP:conf/focs/Luby02} pointed out that one of the potential
applications of LT codes is distributed data storage systems.
Several authors have continued these ideas with fruitful outcomes.
For example,  Plank and Thomason \cite{ldpc} have considered
the practical implementations of LDPC codes for peer-to-peer 
and distributed storage systems and several experimental results
have been reported. In particular, the experimental 
results in \cite{ldpc} show
that LDPC {\em ``codes display their worst performance for $10<n<100$"}
where $n$ is the number of data fragments. Furthermore, 
their experiments show that ``{\em generating good instances of the codes
is a black art...there is an opportunity for theoretical research on
codes for small $n$ to have a very wide-reaching impact}".
Recently, Cao, Yu, Yang, Lou, and Hou \cite{CaoYuYangLouHouLTcodes})
proposed a LT code based secure cloud storage service (LTCS). In 
the LTCS scheme, a data file $F$ is split into $\lambda$ 
packets, each of which is $|F|/\lambda$ bits. 
LT coding process is then used to generate $n\alpha$ encoded packets, 
where $\alpha=\lambda/k\cdot (1+\varepsilon)$. 
These encoded packets are divided into $n$ groups and each of the $n$ 
storage servers receives $\alpha$ packets. One of the basic requirements
from the paper \cite{CaoYuYangLouHouLTcodes} 
is that the original $\lambda$ data packets
should always be recoverable from any $k$ healthy servers. 
For LT codes, there is a small probability that one may not
be able to reconstruct the original data
packets from the $k$ servers. In order to address this challenge, 
the authors in \cite{CaoYuYangLouHouLTcodes}  recommended an
exhaustive search method to divide the encoded symbols
into $n$ groups and check the decodability for 
each $k$ combinations of groups. The process continues 
until one finds out one valid LT coding approach.
This approach is not efficient 
when $k$ and $n$ are relatively larger. Furthermore,
there is no guarantee that the exhaustive search method 
will end with a valid LT code.
The authors did not give any analysis on how efficient
this approach could be or any proof whether that is feasible. 
For the case of $\alpha=1$, the coding scheme 
in \cite{CaoYuYangLouHouLTcodes} is essentially 
a flat XOR code that requires 
a belief propagation (BP) decoder (\cite{DBLP:conf/focs/Luby02}). 
The following example shows that the exhaustive search 
may not succeed for some cases with $\alpha=1$.
\begin{ex}
\label{alpha1ex}
For $n=5$ and $k=3$, the original data 
is divided into three fragments $v_1,v_2,v_3$ and 
coding symbols are stored in $5$ 
servers ($S_1,\cdots, S_5$) such that the original 
data could be recovered from any $3$ servers. 
In order for the belief propagation (BP) decoder to work, 
we must start from an original copy of $v_1$ or $v_2$ or $v_3$. 
Since there could be two erasure faulty servers, three servers 
have to store original copies of the data fragments. 
Without loss of generality, assume that $S_1,S_2$, and $S_3$
store $v_1,v_2$, and $v_3$ respectively.
Again, since there could be two erasure servers,
each data fragment needs to be stored in at least three servers.
Thus both $S_4$ and $S_5$ need to store $v_1\oplus v_2\oplus v_3$. 
Now if both $S_1$ and $S_2$ are faulty, neither $v_1$ nor 
$v_2$ could be recovered.
\hfill$\Box$
\end{ex}

For the case of $\alpha>1$, the coding scheme
in  \cite{CaoYuYangLouHouLTcodes} is a kind of array XOR codes
(not flat XOR codes). In this case,  the robust soliton distribution
will be used to generate the $n\alpha$ encoding symbols. In order 
for the analysis and bounds of the LT code to work, the numbers 
$n,\alpha$, and $\lambda$ in \cite{CaoYuYangLouHouLTcodes}
have to be sufficiently large (this has been confirmed by
the experiments in \cite{ldpc}). For smaller values, these bounds
may not work and the exhaustive search methods in 
\cite{CaoYuYangLouHouLTcodes}  may never end with a successful code.
However, for large enough $n,\alpha$, and $\lambda$, 
the exhaustive search method will be inefficient and may 
be infeasible.

The experiment results from  \cite{ldpc} and the potential challenges
in the scheme \cite{CaoYuYangLouHouLTcodes} show that it is necessary
and important to systematically study the encoding symbol generation 
problems for applying LT code to distributed storage systems.
In the following sections, we will show what we could achieve 
and what we could not achieve with LT codes when applied 
to distributed storage systems.

\section{Flat BP-XOR codes}
\label{ltcodesec}
In this section, we introduce a class of codes called 
flat BP-XOR codes.
In short, flat BP-XOR codes are flat XOR codes that could be
decoded with the Belief Propagation (BP) algorithm for erasure codes.
The BP algorithm for binary symmetric channels is present
in Gallager \cite{gallagerldpc} and is also used 
in artificial intelligence community \cite{pearl}. In our paper, we use 
the BP algorithm for binary erasure channels
(see \cite{Luby:1997:PLC:258533.258573,DBLP:conf/focs/Luby02}).

Let $M=\{0,1\}^l$ be the message symbol set. The length $l$
could be any number and it does not have impact on the coding.
An $[n,k,d]$ flat BP-XOR code is a binary linear code determined 
by a $k\times n$ zero-one valued generator matrix $G$ such that for a given
message vector $x\in M^k$, the corresponding code $y\in M^n$
is computed as $y=xG$ where the addition of two strings in $M$
is defined as the XOR on bits. Furthermore, a flat $[n,k,d]$ BP-XOR 
code requires that if at most $d-1$ components in $y$ are missing,
then $x$ could be recovered from the remaining components of $y$
with the Belief Propagation algorithm,

It is easy to see that each flat BP-XOR code is a
flat XOR code, but the other direction may not hold.
We can consider flat BP-XOR codes as one kind of 
applications of LT codes to reliable storage system design 
with deterministic decoding.

Example \ref{alpha1ex}  shows that flat BP-XOR version of the 
LT code may not be applicable to threshold based distributed storage systems 
as proposed  in \cite{CaoYuYangLouHouLTcodes}. In the following, we 
first mention the folklore fact to support our arguments.

\begin{fact}
\label{singluarlemma}
Let $n\ge k+2$, $k\ge 2$, and $d=n-k+1$. Then there is
no $[n,k,d]$ BP-XOR code.
\end{fact}

The fact could be easily proved by the following observation: 
Let $H=[\beta_1^T, \cdots, \beta_k^T | I_{n-k}]$ be an $(n-k)\times n$ 
parity check matrix. If every $n-k$ columns in the matrix
$[\beta_i^T | I_{n-k}]$ are linearly independent, then $wt(\beta_i)=n-k$,
where $wt(\cdot)$ is the Hamming weight. Thus for $n\ge k+2$, there is neither
binary linear $[n,k,d]$ code nor $[n,k,d]$ BP-XOR code.

Fact \ref{singluarlemma} shows the impossibility 
of designing flat $[n,k,d]$ BP-XOR codes for $n\ge k+2$
and $d=n-k+1$. 
Since flat BP-XOR codes are extremely efficient for encoding and decoding
in practice, we are also interested in flat BP-XOR codes that are not
MDS (maximal distance separable). 
In the following we show theoretical bounds designing 
flat BP-XOR codes for distributed data storage systems. 

For an MDS $[n,k,d]$ code with $d=n-k+1$, we can tolerate $d-1$
erasure faults. The question that we are interested in is: 
for given $n\ge k+2$, what is best distance $d$ we could achieve
for a flat $[n,k,d]$ BP-XOR code?  Fact \ref{singluarlemma}
shows that $d$ must be less than $n-k+1$. 

{\bf Tolerating one erasure fault}: Let $\alpha\in\{1\}^k$.
The generator matrix $\left[I_{k} | \alpha^T\right]$ corresponds to the MDS
flat $[k+1,k,2]$ BP-XOR code that could tolerate 
one erasure fault.

{\bf Tolerating two erasure faults}: Fact \ref{singluarlemma} 
shows that two parity check servers are not sufficient to tolerate
two erasure faults for flat BP-XOR codes. In order to 
tolerate two erasures, we have to consider codes with $n\ge k+3$.
For $n=k+3$, the following generator matrices show the existence
of flat $[5,2,3]$, $[6,3,3]$, and $[7,4,3]$ BP-XOR codes for tolerating 
two erasure faults.
\[
\bigg[
\begin{array}{c|c}
I_2 & 
\begin{array}{ccc}
1 & 0 & 1 \\
0 & 1 & 1
\end{array}
\end{array}
 \bigg]
\left[
\begin{array}{c|c}
I_3 &
\begin{array}{ccc}
0&1&1\\
1&0&1\\
1&1&1
\end{array}
 \end{array}\right]
\left[
 \begin{array}{c|c}
I_4 & 
\begin{array}{ccc}
0&1&1\\
1&0&1\\
1&1&0\\
1&1&1
\end{array}
 \end{array}\right]
\]
Indeed, the above three codes are the only flat
$[k+3,k,3]$ BP-XOR codes tolerating two erasure faults with 
three redundancy columns.
\begin{theorem}
\label{2erasurefaults}
For $n\ge k+3$ and $k\ge 3$, there exists a flat $[n,k,3]$ BP-XOR code if and only if
$k\le 2^{n-k}-(n-k)-1$.
\end{theorem}
{\em Proof}. Let $H=\left[\beta_1^T, \cdots, \beta_k^T|I_{n-k}\right]$ be an $(n-k)\times n$ parity
check matrix.  The code determined by $H$ has minimum distance $3$
if and only if every $2$ columns in $H$ are linearly independent. 
This implies that $H$ is the parity check matrix of a flat $[n,k,3]$ BP-XOR
code if and only if  for every $\beta_i$, we have $wt(\beta_i)\ge 2$
where $wt(\cdot)$ is the Hamming weight.
By the fact that
$$\left|\left\{\beta\in\{0,1\}^{n-k}: wt(\beta)\ge 2
\right\}\right|=2^{n-k}-(n-k)-1,$$
it follows that there exists a flat $[n,k,3]$BP-XOR code if and only if
$k\le 2^{n-k}-(n-k)-1$. 
\hfill$\Box$

{\bf Note}:  It should be noted that the codes we have constructed in
Theorem \ref{2erasurefaults} is the well known Hamming code 
when $k=2^{n-k}-(n-k)-1$. For $k< 2^{n-k}-(n-k)-1$, it is a truncated
version of the Hamming code.

By Theorem \ref{2erasurefaults},  
there is no flat $[k+4,k,3]$ BP-XOR code for $k\ge 12$. Table \ref{table1}
 lists the required redundancy for tolerating two erasure faults
when the value of $k$ changes.
\begin{table}[htb] 
\begin{center}
\caption{Redundancy for flat BP-XOR $[n,k,3]$ codes}
\label{table1}
 \begin{tabular}{ |c|c|c|}
    \hline
 $k$ & required redundancy & BP-XOR code \\ \hline
$2\le k \le 4$ & 3 & $[k+3,k,3]$\\ \hline
$5\le k\le 11$ & 4 & $[k+4,k,3]$\\  \hline
$12\le k\le 26$ &5 & $[k+5,k,3]$\\  \hline
$27\le k\le 57$ &6 & $[k+6,k,3]$\\  \hline
    \end{tabular}
\end{center}
\end{table}
All the codes that we have constructed in Theorem \ref{2erasurefaults}
are systematic. Based on the proof of Theorem \ref{2erasurefaults},
we have the following corollary.
\begin{corollary}
For $n> k$ and $k\ge 3$, there exists an $[n,k,3]$ binary linear code if and only if
there exists a systematic $[n,k,3]$ binary linear code and  if and only if
there exists a systematic flat $[n,k,3]$ BP-XOR code.
\end{corollary}

{\bf Tolerating three erasure faults:} We first prove the following theorem
for the convenience of proving the existence of systematic flat XOR codes.

\begin{theorem}
\label{lineariff}
For $n> k$ and $d\le n-k+1$, there exists an $[n,k,d]$ 
binary linear code code if and only if there exists an $(n-k)\times k$ 
matrix $A=(\beta_1^T, \cdots, \beta_k^T)$ with the following properties:
\begin{enumerate}
\item $\beta_i\in \{0,1\}^{n-k}$ for $1\le i\le k$
\item\label{cond2} Let $d_1+d_2=d-1$. If we remove $d_2$ rows from $A$,
then every $d_1$ columns of the remaining matrix are linearly independent.
\end{enumerate}
\end{theorem}

{\em Proof}. First, it is straightforward to show that 
the condition \ref{cond2} in the Theorem implies that 
$wt(\beta_i)\ge d-1$ for $1\le i\le k$, where $wt(\cdot)$ 
is the Hamming weight. It is also straightforward to show
that the condition \ref{cond2} in the Theorem implies that 
every $d-1$ columns in the matrix $[A|I_{n-k}]$ are linearly
independent. Thus, the linear code corresponding to the parity check matrix 
$[A|I_{n-k}]$ is a binary linear $[n,k,d]$ code. Note that the
generator matrix corresponding to the parity check matrix 
$[A|I_{n-k}]$ is $G=[I_k | A^T]$.

For the other direction, assume that there exists a $k\times n$ 
generator matrix $G$ for an $[n,k,d]$ binary linear code.
Let $\alpha_1, \cdots, \alpha_n$ be the $n$ columns 
of $G$. Without loss of generality, 
we may assume that $\alpha_1, \cdots, \alpha_k$ are 
linearly independent. We may also assume that 
$$
\left(
\begin{array}{l}
\alpha_1^T\\
\vdots \\
\alpha_n^T
\end{array}\right)=
\left(
\begin{array}{c}
I_k\\
A
\end{array}
\right)
\left(
\begin{array}{l}
\alpha_1^T\\
\vdots \\
\alpha_k^T
\end{array}\right)
$$
where $A=(\beta_1^T, \cdots, \beta_k^T)$ and 
$\beta_i\in\{0,1\}^{n-k}$.

Since the code has the minimum distance $d$, the remaining 
generator matrix $G$ should have a rank of $k$ after removing 
any $d-1$ columns from $G$. 
Let $d_1+d_2=d-1$ and assume that we remove $d_1$ columns 
$\alpha_{i_1}, \cdots, \alpha_{i_{d_1}}$ for $i_u\le k$ and $d_2$ 
columns $\alpha_{k+j_1}$, $\cdots$, $\alpha_{k+j_{d_2}}$ for $j_u\le n-k$
from the generator matrix $G$. 
Then $\alpha_{i_1}, \cdots, \alpha_{i_{d_1}}$ should be able to be 
linearly generated from the columns $\alpha_i$ for $i\ge k+1$
and $i\not=j_1,\cdots, j_{d_2}$. This is equivalent to the
requirements that the rows $i_1$, $\cdots$, $i_{d_1}$ of
$I_k$ could be linearly generated from the remaining rows of $A$
after removing the rows $j_1$, $\cdots$, $j_{d_2}$ from $A$.
It follows that the remaining columns $i_1$, $\cdots$, $i_{d_1}$ 
of $A$ are linearly independent after removing the rows 
$j_1$, $\cdots$, $j_{d_2}$ from $A$. This completes the proof
of the Theorem. \hfill$\Box$

By Theorem \ref{lineariff}, we have the following results. 
\begin{theorem}
\label{3erasurefaults}
For $n\ge k+4$, there exists a systematic flat XOR $[n,k,4]$ code if 
and only if 
$$k\le \left\{
\begin{array}{ll}
2^{n-k-1}-n+k& \mbox{ if } n-k\mbox{ is even}\\
2^{n-k-1}-n+k-1& \mbox{ if } n-k\mbox{ is odd}
\end{array} \right.$$
\end{theorem}
{\em Proof}. 
Let $$X=\{\beta: \beta\in\{0,1\}^{n-k}, wt(\beta)=3, 5, 7,\cdots\}.$$
Then 
\begingroup
\def\arraystretch{1.5}
$$\begin{array}{lll}
|X|&=& \displaystyle\sum_{i\ge 3, i\mbox{ is odd }}{n-k \choose i}\\
&=&\displaystyle\sum_{i\ge 3, i\mbox{ is odd }}\left(
{n-k-1 \choose i-1}+{n-k-1 \choose i}
\right)\\
&=&\left\{
\begin{array}{ll}
2^{n-k-1}-n+k& \mbox{ if } n-k\mbox{ is even}\\
2^{n-k-1}-n+k-1& \mbox{ if } n-k\mbox{ is odd}
\end{array} \right.
\end{array}$$
\endgroup
Define an $(n-k)\times k$ matrix 
$A=(\beta_1^T, \cdots, \beta_k^T)$ where $\beta_i\in X$.
It is straightforward to show that this matrix $A$ satisfies the condition 
2 of Theorem \ref{lineariff} for $d=4$ (alternatively, every three columns
in the parity check matrix $[A|I_{n-k}]$ are linearly independent). 
Thus the binary linear code corresponding 
to the parity check matrix $[A|I_{n-k}]$ (or the generator matrix $[I_k|A^T]$)
is a flat XOR $[n,k,4]$ code.

For the other direction, it suffices to show that $X$ is a maximal set 
that satisfies the condition 2 of Theorem \ref{lineariff} for $d=4$. This is
proved by observing the fact that every even Hamming weight vector
$\beta\in\{0,1\}^{n-k}$ is equal to $\beta_1+\beta_2$ for some $\beta_1,\beta_2\in X$.
This completes the proof of the theorem.
\hfill$\Box$

In Theorem \ref{3erasurefaults}, we established a necessary and 
sufficient condition for designing systematic flat XOR codes tolerating 
three erasure faults. However, the codes 
we constructed in Theorem \ref{3erasurefaults} are not necessarily 
flat BP-XOR codes. For example, let  $n=7, k=3$, 
$d=4$, and 
$\beta_1=(1,1,1,0)$,  
$\beta_2=(0,1,1,1)$, and 
$\beta_3=(1,0,1,1)$. Then the corresponding code has the following generator
matrix:
$$\left[
 \begin{array}{c|c}
I_3 & 
\begin{array}{cccc}
1&1&1&0\\
0&1&1&1\\
1&0&1&1
\end{array}
 \end{array}\right]$$
It is straightforward that this is not a flat BP-XOR code since if we remove the first
three columns from the above generator matrix, no column in the remaining generator
matrix has Hamming weight $1$. Indeed, it is easy to show that for  
 $n=7, k=3$,  and $d=4$, there is no flat $[7,3,4]$ BP-XOR code.
The reason is that in order for a $[7,3,4]$ linear code to be a flat BP-XOR code,
we have to have four columns with Hamming weight $1$ in the generator matrix.
Furthermore, we need to have Hamming weight $4$ for each row.
Without loss of generality, we may assume that the column $(1,0,0)^T$ occurs
twice in the generator matrix. Then we have to have  three columns in the generator matrix 
with the format $(b, 1,1)^T$ where
$b=0,1$. In another word, two columns of the generator matrix are identical,
which will reduce the code distance to $3$.

The above discussion shows that the condition in Theorem \ref{3erasurefaults}
is not valid for the existence of flat BP-XOR code tolerating three erasure faults.
Though it is interesting to identify necessary and sufficient conditions 
for the existence of BP-XOR codes tolerating three or more erasure faults, 
it is sufficient for us to use the flat XOR codes in distributed storage systems
since a simple XOR based Gauss elimination methods could be used to 
recover the original data content in front of erasure faults. 
This observation tells us that LT code (i.e., the flat BP-XOR code) 
may not be the best choices for distributed storage systems in some cases.

As an example, Table \ref{table2} lists the required redundancy for tolerating 
three erasure faults when the value of $k$ changes.
\begin{table}[htb] 
\begin{center}
\caption{Redundancy for flat XOR $[n,k,4]$ codes}
\label{table2}
 \begin{tabular}{ |c|c|c|}
    \hline
 $k$ & required redundancy & flat XOR code \\ \hline
$2\le k \le 4$ & 4 & $[k+4,k,4]$\\ \hline
$5\le k\le 10$ & 5 & $[k+5,k,4]$\\  \hline
$11\le k\le 26$ &6 & $[k+6,k,4]$\\  \hline
$27\le k\le 56$ &7 & $[k+7,k,4]$\\  \hline
    \end{tabular}
\end{center}
\end{table}

{\bf Tolerating four or more erasure faults:} 
In general, we are also interested in designing flat BP-XOR codes 
for tolerating more than three erasure faults. 
For distributed storage systems we could generally use 
nested techniques (e.g., the similar techniques as nested RAID array).
In  the following, we present several sufficient conditions 
for tolerating four or more erasure faults. Normally these conditions
are not necessary. 
It should be noted that for general binary linear codes,
there are well known bounds (see, e.g., 
Verhoeff \cite{Verhoeff:1987:UTM:41075.41082}). However, the codes 
corresponding to these bounds are not necessarily flat BP-XOR codes.

\begin{theorem}
\label{4erasurefaults}
For $n\ge k+5$, there exists a systematic flat XOR $[n,k,5]$ code if 
$k$ is less than
$$
\begin{array}{l}
\left\lfloor \displaystyle\frac{n-k-2}{2}\right\rfloor+
2\left\lfloor \left(\left\lceil \displaystyle\frac{n-k}{2}\right\rceil-2\right)/2\right\rfloor\\
\quad\quad\quad\quad\quad\quad\quad\quad\quad\quad
+ 2\left\lfloor \left(\left\lceil \displaystyle\frac{n-k}{4}\right\rceil-2\right)/2\right\rfloor.
\end{array}$$
\end{theorem}
{\em Proof}. 
Let $U=\{a_1,\cdots, a_{n-k}\}$ be an $n-k$ element set. 
In the following, we construct four-element subsets 
of $U$ so that the characteristic sequences
of these subsets could be used as the columns of the parity check
matrix. It will be convenient for the reader to understand the following
subset definitions if the elements of $U$ are interpreted as leaf nodes 
on a binary tree of depth $\lfloor \log_2 (n-k)\rfloor$.
\begingroup
\def\arraystretch{1.5}
$$\begin{array}{l}
V_i^1=\{a_1,a_2, a_{2i+1}, a_{2i+2}\}\mbox{ for }1\le i\le \left\lfloor \frac{n-k-2}{2}\right\rfloor,\\
V_i^{2,0}=\{a_1,a_3, a_{4i+1}, a_{4i+3}\}\mbox{ for }1\le i\le 
\left\lfloor \left(\lceil \frac{n-k}{2}\rceil-2\right)/2\right\rfloor,\\
V_i^{2,1}=\{a_2,a_4, a_{4i+2}, a_{4i+3}\}\mbox{ for }1\le i\le 
\left\lfloor \left(\lceil \frac{n-k}{2}\rceil-2\right)/2\right\rfloor,\\
V_i^{3,0}=\{a_1,a_5, a_{8i+1}, a_{8i+5}\}\mbox{ for }1\le i\le 
\left\lfloor \left(\lceil \frac{n-k}{4}\rceil-2\right)/2\right\rfloor,\\
V_i^{3,1}=\{a_4,a_8, a_{4i+2}, a_{8i+5}\}\mbox{ for }1\le i\le 
\left\lfloor \left(\lceil \frac{n-k}{4}\rceil-2\right)/2\right\rfloor,\\
\cdots \cdots
\end{array}$$
\endgroup
Let $\beta_1, \cdots, \beta_w$ be the characteristic sequences of the above sets.
Then it is straightforward that the parity check matrix 
$H=[\beta_1^T,\cdots, \beta_w^T |I_{n-k}]$ corresponds to a systematic flat XOR code
of minimum distance $5$.  The code has distance $5$ since every $4$ columns in $H$
are linearly independent by the facts that 
(1)  for any $\beta_1,\beta_2$, we have $wt(\beta_1+\beta_2)>2$; and 
(2) any three or four $\beta$ are linearly independent.  The two facts follow from the
construction. This completes the Proof of the Theorem.
\hfill$\Box$

As an example, Table \ref{table3} lists the required redundancy of flat XOR codes
for tolerating $4$ erasure faults based on Theorem \ref{4erasurefaults}.
\begin{table}[htb] 
\begin{center}
\caption{Redundancy for flat XOR $[n,k,5]$ codes}
\label{table3}
 \begin{tabular}{ |c|c|c|}
    \hline
$k$ & required redundancy & flat XOR code \\ \hline
$k \le 2 $ & 6 & $[k+6,k,5]$\\ \hline
$3\le k \le 4$ & 7 & $[k+7,k,5]$\\ \hline
$5\le k \le 9$ & 8 & $[k+8,k,5]$\\  \hline
    \end{tabular}
\end{center}
\end{table}
Though the conditions in Theorem \ref{4erasurefaults} are not necessary in general.
The bounds in Table \ref{table3} matches the bounds for general binary linear codes
(see \cite{Verhoeff:1987:UTM:41075.41082}). Thus the conditions  
in Theorem \ref{4erasurefaults}  are also necessary for $k\le 9$.

\section{Array BP-XOR codes for distributed storage systems}
\label{arraybpxorsec}
Array codes have been studied  extensively for burst error 
correction in communication systems and storage systems
(see, e.g., 
\cite{DBLP:journals/tc/BlaumBBM95,Blaum96mdsarray,DBLP:journals/tit/BlaumR93,Blaum99onlowest-density,Cassuto:2009:CLD:1669446.1669469,Xu98lowdensity,Xu99x-code:mds}).
Array codes are linear codes where information and parity data are placed
in a two dimensional matrix array. Appropriately designed array codes 
such as EVENODD \cite{DBLP:journals/tc/BlaumBBM95}, RDP \cite{DBLP:conf/fast/CorbettEGGKLS04},
STAR \cite{Huang05anefficient}, X-code \cite{Xu99x-code:mds}
are very useful for high speed storage application systems 
since they enjoy low-complexity decoding and low update complexity.
Most of these array codes are designed for RAID array based storage
systems with the specific requirements such as systematic code, efficient decoding 
algorithm, and minimum update complexity, where update complexity refers to the 
number of encoding data symbols that need to be updated if one information data symbol is changed.

For distributed storage systems such as cloud storage, we may not need 
the code to be systematic. 
As studied in \cite{DBLP:conf/focs/Luby02,ldpc,CaoYuYangLouHouLTcodes},
LT code or digital fountain techniques could be a better choice for 
distributed storage systems. However, as we have mentioned in previous 
sections and as supported by the experimental results in \cite{ldpc},
the probabilistic bounds in LT code performs well only asymptotically 
when the number $n$ of encoding symbols increase. For small numbers of 
$n<100$, the codes display their worst performance. 
 Thus it is important to study the applicable coding schemes
with better performance for distributed storage systems. As we have noticed, 
one of the major advantages that contribute to the efficiency of LT decoding
process is the Belief Propagation (BP) process. In the following, we design
a kind of array codes that could be efficiently decoded using the BP-process.
We will call such kind of codes array BP-XOR codes. Appropriately designed
array BP-XOR codes could achieve the MDS property from both
communication and storage aspects: for $k$ blocks of the original data,
surviving storage servers only need to 
store $k$ blocks of encoding blocks. Note that in LT codes, 
in order to  decode $k$ blocks of data with probability $1-\delta$, 
$k+O(\sqrt{k}\ln^2(k/\delta))$ blocks of encoding blocks are needed.

Array BP-XOR code is defined as follows. Let $\alpha_1, \cdots, \alpha_k
\in \{0,1\}^l$ be the data fragments that we want to encode, where $l$ is
any fixed number. A $t$-erasure tolerating array BP-XOR code 
is an $m\times n$ matrix 
${\cal C}=[\sigma_{i,j}]_{1\le i\le m, 1\le j\le n}$  
such that: 
\begin{enumerate}
\item Each $\sigma_{i,j}$ is the XOR of one or more
elements from the data fragments $\alpha_1, \cdots, \alpha_k$. 
\item $\alpha_1, \cdots, \alpha_k$ could be recovered 
from any $n-t$ columns of the matrix using the binary erasure 
channel based BP algorithm.
\end{enumerate}

If we add the restriction that each element in
${\cal C}=[\sigma_{i,j}]_{1\le i\le m, 1\le j\le n}$  
be the XOR of at most two elements from the data fragments 
$\alpha_1, \cdots, \alpha_k$, then the restricted 
array BP-XOR codes are equivalent to the edge-colored graph models 
introduced by Wang and Desmedt in \cite{Wang:2011:EGA:1975023.1975459} 
for tolerating network homogeneous faults.

\subsection{Edge-colored graphs}
In this section, we first describe the edge-colored graph model by
Wang and Desmedt  \cite{Wang:2011:EGA:1975023.1975459}. The reader should
be reminded that the edge-colored graph model in   
\cite{Wang:2011:EGA:1975023.1975459} is slightly different 
from the edge-colored graph definition in most literatures.
In most literatures, the coloring of the edges is required to meet the condition
that no two adjacent edges have the same color. This condition
is not required in the definition of \cite{Wang:2011:EGA:1975023.1975459}.

\begin{definition}
(Wang and Desmedt  \cite{Wang:2011:EGA:1975023.1975459}) 
An edge-colored graph is a tuple $G(V,E,C,f)$, with $V$ the node set,
$E$ the edge set, $C$ the color set, and $f$ a map from $E$ onto $C$.
The structure
$$
{\cal Z}_{C,t}=\{Z: Z\subseteq E\,\, \hbox{and}\,\,\,
|f(Z)|\leq t\}.
$$
is called a $t$-{\it color\/} adversary structure.
Let $A,B \in V$ be distinct nodes of $G$.
$A,B$ are called $(t+1)$-{\em color connected\/} for $t\ge 1$ if 
for any color set $C_{t}\subseteq C$ of size $t$, 
there is a path $p$ from $A$ to $B$ 
in $G$ such that the edges on $p$ do not contain any color in $C_{t}$.
An  edge-colored graph $G$ is  $(t+1)$-{\em color connected} if and only
if for any two nodes $A$ and $B$ in $G$, they are 
$(t+1)$-color connected.
\end{definition}

\begin{figure}[htb]
\centering
\subfigure[$G_{4,1}$]{
\centering{\includegraphics[scale=0.9]{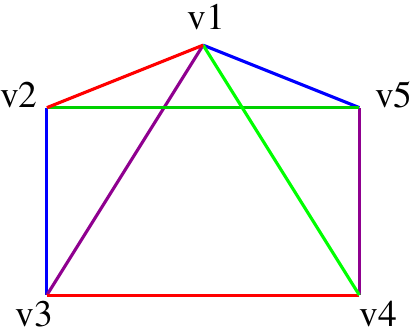}}	
\label{figm4t2}
}
\subfigure[$G_{4,2}$]{
	\label{fign7m4t2}
\centering{\includegraphics[scale=0.9]{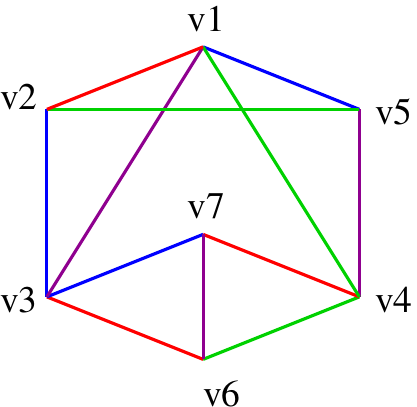}}
}
\caption{$3$-color 
connected edge-colored graphs}
\label{3connectedgraphs}
\end{figure}

As an example, Figure \ref{3connectedgraphs} contains
two 3-color connected  edge-colored graphs $G_{4,1}$ and $G_{4,2}$.
$G_{4,1}$  contains $5$ nodes, $8$ edges, 
and $4$ colors. $G_{4,2}$ contains $7$ nodes, $12$ edges, and $4$ colors.

A general $3$-color connected edge-colored graph with $k$ nodes 
can be constructed as follows. 
\begin{enumerate}
\item For $k=4r+1$,  the $v_1$ node of $r$ copies of  $G_{4,1}$ 
are glued together to form a $3$-color connected edge-colored 
graph $G$ with $4r+1$ nodes and $8r$ edges. 
\item For $k=4r+3$,  the $v_1$ node of $r-1$ copies of  $G_{4,1}$ 
and one copy of $G_{4,2}$ are glued together to form a $3$-color connected 
edge-colored graph $G$ with $4r+3$ nodes, $8r+4$ edges. 
\item For $k=4r+2$ (respectively $k=4r+4$) with $r\ge 1$, one node is added
to the  $3$-color connected edge-colored graph $G$ with $k=4r+1$ nodes
(respectively $k=4r+3$ nodes) by connecting this node to any $3$ 
nodes within the graph with distinct colors. 
The resulting graph is a $4$-color 
connected edge-colored graph with $4r+2$ (respectively $4r+4$) 
nodes and $8r+3$ (respectively $8r+7$) edges. 
\end{enumerate}

For convenience, an edge-colored graph could also be represented by
a table, where the edges with same colors are put in the same column.
For example,  $G_{4,1}$ and $G_{4,2}$ in Figure \ref{3connectedgraphs} 
are represented in Table \ref{g41g42table}.
\begin{table}[htb] 
\begin{center}
\caption{Table representation of edge-colored graphs $G_{4,1}$ and $G_{4,2}$}
\label{g41g42table}
 \begin{tabular}{ |c|c|c|c|c|}   \hline
\multirow{2}{*}{ $G_{4,1}$ } &$\langle v_1, v_2 \rangle$ & $\langle v_2, v_3 \rangle$ & $\langle v_4, v_5 \rangle$ & $\langle v_2, v_5 \rangle$\\ 
&$\langle v_3, v_4 \rangle$ & $\langle v_1, v_5 \rangle$ & $\langle v_1, v_3 \rangle$ & $\langle v_1, v_4 \rangle$\\ \hline
\multirow{3}{*}{ $G_{4,2}$ } & $\langle v_1, v_2 \rangle$ & $\langle v_2, v_3 \rangle$ & $\langle v_4, v_5 \rangle$ & $\langle v_2, v_5 \rangle$\\ 
& $\langle v_3, v_6 \rangle$ & $\langle v_1, v_5 \rangle$ & $\langle v_6, v_7 \rangle$ & $\langle v_1, v_4 \rangle$\\ 
& $\langle v_4, v_7 \rangle$ & $\langle v_3, v_7 \rangle$ & $\langle v_1, v_3 \rangle$ & $\langle v_4, v_6 \rangle$\\ \hline
   \end{tabular}
\end{center}
\end{table}

Wang and Desmedt \cite{Wang:2011:EGA:1975023.1975459} showed several
constructions of edge-colored graphs with certain color connectivity. In the following,
we present a general construction of $(t+1)$-color connected edge-colored graphs
using perfect one-factorizations of complete graphs. We use $K_n=(V,E)$ to denote
the complete graph with $n$ nodes. For an even $n$, a one-factor of $K_n$ is a set
of pairwise disjoint edges that partition the set of nodes in $V$. A one-factorization
of $K_n$ ($n$ is even) is a set of one-factors that partition the set of edges $E$.
A one-factorization is called perfect if the union of every two distinct one-factors
is a Hamiltonian circuit. It is shown 
(see, e.g., Anderson \cite{AndersonTopology} and 
Kobayashi \cite{DBLP:journals/gc/Kobayashi89} )
that perfect one-factorizations for 
$K_{p+1}$, $K_{2p}$, and certain $K_{2n}$ do exist, where $p$ is a prime number.

\begin{theorem}
\label{p1ftocolor}
Let $n$ be an even number such that there is a perfect one-factorization 
$F_1, \cdots, F_{n-1}$ for $K_n$.  
For each $t\le n-3$, there exists a $(t+1)$-color connected 
edge-colored graph $G$ with $n-1$ nodes, $(t+2)(n/2-1)$ edges,  and $t+2$ colors.
\end{theorem}
{\em Proof}. Let $V=\{v_1, \cdots, v_{n-1}\}$,
$F'_i=F_i\setminus \{\langle v_n, v\rangle\}$,
and $E=F_1'\cup \cdots \cup F'_{t+2}$ and 
color all edges in $F'_i$ with color $c_i$ for $i\le t+2$. Then it is straightforward
to check that the edge-colored graph $(V,E)$ is $(t+1)$-color connected,
$|V|=n-1$,  and $|E|=(t+2)(n/2-1)$.  \hfill$\Box$

{\em Remarks on Proof of Theorem \ref{p1ftocolor}}: 
Since only node connectivity instead of Hamiltonian 
circuit is required for $(t+1)$-color connected graphs, we could 
use $F'_i$ instead of $F_i$ to construct the edge-colored graphs.
By using $F'_i$, we reduce $t+2$ edges and one node in the resulting 
edge-colored graph. This will help us to keep the minimum cost for connectivity.

\subsection{Constructing array BP-XOR codes from edge-colored graphs}
We now use edge-colored graphs to construct array BP-XOR codes. As an example,
we first give the BP-XOR code corresponding to the graph $G_{4,2}$ in 
Table \ref{g41g42table}. As Step 1, the $G_{4,2}$ part in Table \ref{g41g42table}
is converted to the code in Table \ref{g42code1}.
\begin{table}[htb] 
\begin{center}
\caption{First step code for  $G_{4,2}$}
\label{g42code1}
 \begin{tabular}{ |c|c|c|c|}   \hline
 $v_1\oplus v_2$ & $v_2\oplus v_3 $ & $v_4\oplus  v_5$ & $ v_2\oplus  v_5$\\ \hline
$v_3\oplus  v_6$ & $v_1\oplus  v_5$ & $v_6\oplus v_7$ & $v_1\oplus  v_4$\\ \hline
$v_4\oplus  v_7$ & $v_3\oplus v_7$ & $v_1\oplus v_3$ & $v_4\oplus v_6 $\\ \hline
   \end{tabular}
\end{center}
\end{table}
In the step 2, choose any fixed node and remove all of its occurrence from the 
code in Table \ref{g42code1}. For convenience, we choose to remove the occurrence 
of $v_7$ and get the BP-XOR code in Table \ref{g42code2}.
\begin{table}[htb] 
\begin{center}
\caption{BP-XOR code corresponding to $G_{4,2}$}
\label{g42code2}
 \begin{tabular}{ |c|c|c|c|}   \hline
 $v_1\oplus v_2$ & $v_2\oplus v_3 $ & $v_4\oplus  v_5$ & $ v_2\oplus  v_5$\\ \hline
$v_3\oplus  v_6$ & $v_1\oplus  v_5$ & $v_6$ & $v_1\oplus  v_4$\\ \hline
$v_4$ & $v_3$ & $v_1\oplus v_3$ & $v_4\oplus v_6 $\\ \hline
   \end{tabular}
\end{center}
\end{table}

It is easy to check that the data fragments $v_1,\cdots, v_6$ can be
recovered from any two columns of coding symbols. It is also straightforward
to observe that the code in Table  \ref{g42code2} achieves optimal space
and communication bandwidth in the event of two column erasures.

In the following, we give the general construction of BP-XOR code from 
edge-colored graphs.
Let $v_1, v_2, \cdots, v_k\in\{0,1\}^l$ be data blocks that we want to encode, 
where $l$ is any fixed length.  Let  $G(V,E,C,f)$ be a $(t+1)$-color connected
edge-colored graph with $V=\{v_1, \cdots, v_k, v_{k+1}\}$, $|E|=\lambda$, and 
$C=\{c_1, c_2, \cdots, c_n\}$.
If we consider the nodes in  the edge-colored graph $G(V,E,C,f)$
as data blocks, edges as their parity check blocks of the adjacent 
nodes, and colors on the edges as labels for placing the parity checks
into different columns of the array codes, then following steps construct
an $m\times n$ array BP-XOR codes, where 
$m=\max_{c\in C}\{|Z|: Z\subseteq E,  f(Z)=c\}$.
\begin{enumerate}
\item For $1\le i\le n$, let 
$$\beta'_i=\{v_{i}\oplus v_{j}: \langle v_{i},v_{j}\rangle\in E,
                f(\langle v_{i},v_{j}\rangle)=c_i\} .$$
\item For each $\beta_i'$, replace the entry $v_{k+1}\oplus v$ with $v$ if such entry exists.
Furthermore, if $|\beta_i'|$ is smaller than $m$, add empty element
to $\beta'_i$ to make it an $m$-length vector $\beta_i$. 
\item The array BP-XOR code is then specified by the $m\times n$ matrix
${\cal C}_G=(\beta_1^T, \cdots, \beta^T_n)$.
\end{enumerate}

Next we show that the above defined array BP-XOR code ${\cal C}_G$
can tolerate $t$ column erasure faults. Let $C_t\subset C$ be any 
set of $t$ colors of the graph $G$ and assume that $t$ columns 
corresponding to the color set $C_t$ are missing in ${\cal C}_G$. 
Since the graph $G$ is $(t+1)$-color connected, for any node 
$v_{i_0}\in V$, we have a path 
$p=\langle v_{k+1}, v_{i_1}, v_{i_2}, \cdots, v_{i_{j}}, v_{i_0}\rangle$ 
without using any colors in $C_t$. Thus $v_{i_0}$ could be recovered 
by the following equation
$$v_{i_0}= v_{i_1}\oplus  (v_{i_1}\oplus v_{i_2}) \oplus \cdots\oplus (v_{i_j}\oplus v_{i_0})$$
where 
$v_{i_1}, v_{i_1}\oplus v_{i_2}$, $\cdots$, $v_{i_j}\oplus v_{i_0}$ are all contained in the 
non-missing columns. Thus the Belief Propagation process could be used 
to recover the entire data blocks $v_1, \cdots, v_k$ from the non-missing
columns with only $k$ XOR operations on the encoding symbols. 

\subsection{Constructing edge-colored graphs from array BP-XOR codes}
In this section, we show that for each array BP-XOR code, we could
construct a corresponding edge-colored graph. 
\begin{theorem}
\label{graphtobpxorthm}
Let ${\cal C}$
be an $m\times n$  array BP-XOR code with the following properties:
\begin{enumerate}
\item ${\cal C}$ is $t$-erasure tolerating;
\item ${\cal C}$ contains $k$ information symbols; and
\item ${\cal C}$ contains only degree one and two encoding symbols.
\end{enumerate}
Then there exists a $(t+1)$-color connected edge-colored graph
$G(V,E,C,f)$ with $|V|=k+1$, $|E|=mn$,  and $|C|=n$.
\end{theorem}

{\it Proof.}  Let $v_1, \cdots, v_k$ be the information symbols of 
${\cal C}=[a_{i,j}]_{(i,j)\in [1,m]\times [1,n]}$
and $v_{i_1}, \cdots, v_{i_u}$ be a list of degree one encoding symbols in ${\cal C}$.
Then the $(t+1)$-color connected edge-colored graph $G(V,E,C,f)$ is defined by the
following steps:
\begin{enumerate}
\item $V=\{v_1, \cdots, v_k, v_{k+1}\}$;
\item $E=\displaystyle\cup_{j\in [1, u]}\{\langle v_{k+1}, v_{i_j}\rangle\}
\cup\{\langle v_{i'},v_{j'}\rangle: a_{i,j}=v_{i'}\oplus v_{j'}\in {\cal C}\}$;
\item $C=\{c_1, \cdots, c_n\}$;
\item for each $a_{i,j}=v_{i'}\oplus v_{j'}\in {\cal C}$, let $f(\langle v_{i'},v_{j'}\rangle)=c_j$ and for 
each $a_{i,j}=v_{i'}\in {\cal C}$ let $f(\langle v_{k+1},v_{i'}\rangle)=c_j$
\end{enumerate}
Let $C_t$ be a color set of size $t$ and  $v_i$ and $v_j$ be two nodes.
Since the code ${\cal C}$ is $t$-erasure tolerating, both $v_i$ and $v_j$
could be recovered from encoding symbols not  contained 
in the columns corresponding to the colors in $C_t$. Thus 
there exists a path $p$ ($q$ respectively) connecting $v_{k+1}$ to $v_i$
(to $v_j$ respectively) without using $C_t$-colored edges.
It follows that $G(V,E,C,f)$ is $(t+1)$-color connected.
\hfill{$\Box$}

\section{Examples of bandwidth optimal array BP-XOR codes}
\label{arraybpxorexamplesec}
In this section, we use edge-colored graphs in
Theorem \ref{p1ftocolor} to construct $m\times n$
BP-XOR codes that could tolerate $n-2$  erasure columns.
The general process is as follows: For a given number $n$ 
of code columns and a number $t$ of erasure
columns, we first design $(t+1)$-color connected edge-colored
graphs with $n$ colors and the smallest number of graph edges.
The resulting edge-colored graph is then converted to the BP-XOR 
code with the process described in the previous section.

In order to design an $m\times n$ BP-XOR code tolerating
$n-2$ erasure columns, find the smallest $p$ (or $2p$) such 
that $n\le p$ (or $n\le 2p-1$), where $p$ is an odd prime.
Suppose $p$ is such a prime with $n\le p$. Then Table \ref{pcolorgraph}
defines a $(p-1)$-color connected edge-colored graphs with $p$ nodes
and $p$ colors (based on the 
perfect one-factorization of $K_{p+1}$ in \cite{DBLP:journals/gc/Kobayashi89}).
\begin{table}[htb] 
\begin{center}
\caption{$(p-1)$-color connected edge-colored graphs}
\label{pcolorgraph}
 \begin{tabular}{ |c|c|c|}   \hline
$\langle v_1, v_{p-1}\rangle$ & $\cdots$ & $\langle v_p, v_{p-2}\rangle$ 
\\ \hline
$\langle v_2, v_{p-2}\rangle$ & $\cdots$ & $\langle v_1, v_{p-3}\rangle$
\\ \hline
$\cdots$ & $\cdots$ & $\cdots$
\\ \hline
$\langle v_{(p-1)/2}, v_{(p+1)/2}\rangle$ & $\cdots$ &  $\langle v_{(p-3)/2}, v_{(p-1)/2}\rangle$ 
\\ \hline
   \end{tabular}
\end{center}
\end{table}
In Table \ref{pcolorgraph}, if we consider  the first column
as a sequence of numbers: 
$1, p-1; 2, p-2; \cdots; (p-1)/2, (p+1)/2,$
then the $i$th column of the table is defined by the following 
sequence of numbers (operations are mod $p$ and $0$ is replaced with $p$):
$$1+i, p-1+i; 2+i, p-2+i; \cdots; (p-1)/2+i, (p+1)/2+i.$$
The above edge-colored graph is then converted to the 
$(p-1)/2\times p$ BP-XOR code in Table \ref{pcolorgraphcode}
where $m=(p-1)/2$.
\begin{table}[htb] 
\begin{center}
\caption{$(p-1)/2\times p$ BP-XOR code}
\label{pcolorgraphcode}
 \begin{tabular}{ |c|c|c|c|}   \hline
$v_1\oplus v_{p-1}$ & $\cdots$ & $v_{p-1}\oplus v_{p-3}$  & $v_{p-2}$ 
\\ \hline
$v_2\oplus v_{p-2}$ & $\cdots$ & $v_{p-4}$ &$v_1\oplus v_{p-3}$
\\ \hline
$\cdots$ & $\cdots$ & $\cdots$ & $\cdots$
\\ \hline
$v_m\oplus v_{m+1}$ & $\cdots$ &$v_{m-2}\oplus v_{m-1}$   &$v_{m-1}\oplus v_{m}$ 
\\ \hline
   \end{tabular}
\end{center}
\end{table}
Then an $m\times n$ BP-XOR code is obtained by taking any of the 
$n$ columns in Table \ref{pcolorgraphcode}. 
Since the edge-colored graph in Table \ref{pcolorgraph} is 
$(p-1)$-color connected, it follows that the above constructed 
$m\times n$ BP-XOR code could tolerate $n-2$ erasure columns.
In another word, the original data content $F$ is divided into 
$p-1$ fragments $v_1, \cdots, v_{p-1}\in \{0,1\}^l$ of equal length ($l$ bits) 
and  are stored in $n$ servers according to the BP-XOR codes, then
the original data content $F$ could always be recovered from any
two surviving servers. It should also be noted that 
each storage server stores $(p-1)l/2$ bits of data and
the total data stored at two storage servers are $(p-1)l=|F|$ bits. 
Thus the BP-XOR code is optimal in bandwidth and space.

We should also note that the $(p-1)/2\times p$ BP-XOR code in
Table \ref{pcolorgraphcode} is equivalent to the code designed by  
Zaitsev, Zinov'ev, and Semakov \cite{ZaiZinSem83} 
which was reformulated later as the dual code of B-code in 
\cite{Xu98lowdensity} using perfect 
one-factorization of complete graphs.

\section{The limitation of degree two encoding symbols}
\label{comparisonsec}
In this section we analyze the limitation of array BP-XOR codes
when only degree one and two encoding symbols are allowed.
Using the results for array BP-XOR codes, we will get new results
for edge-colored graph models.

For a $t$-erasure tolerating array BP-XOR code of size $m\times n$,
we could achieve space and bandwidth optimal property
if there are $k=(n-t)m$ information symbols of same length.
The following theorem provides a necessary condition for the 
existence of array BP-XOR codes when only degree one and two
encoding symbols are used.

\begin{theorem}
\label{necessaryarraybpxor}
Let ${\cal C}=[a_{i,j}]_{(i,j)\in [1, m]\times [1, n]}$ be a $t$-erasure
tolerating array BP-XOR code with $k=(n-t)m$ information symbols and 
${\cal C}$ only use degree one and two encoding symbols. Assume that
$n_0=n-t>2$, then we have 
$$n\le \frac{n_0-1}{n_0-2}\left(
n_0-\frac{2}{(n_0-2)m+1}
\right).$$
\end{theorem}

{\em Proof.} 
By the fact that ${\cal C}$ is $t$-erasure tolerating, 
each information symbol must occur 
in at least $t+1$ columns. 
Since there are $n_0m$ information symbols (data fragments) 
to encode, the total number of information symbol occurrences 
in ${\cal C}$ is at least $n_0m(t+1)$.

In order for the BP decoding process to work, we must start from
a degree one encoding symbol. Thus we need to have 
at least $t+1$ degree one encoding symbols in distinct columns
of ${\cal C}$. This implies that we could use at most 
$mn-(t+1)$ cells to hold encoding symbols for degree
two. In another word,  ${\cal C}$ contains
at most $2 (mn-(t+1))+t+1$ occurrences of 
information symbols. By the above fact, we must have
$$n_0m(t+1) \le 2 (mn-(t+1))+t+1.$$
By rearranging the terms, we get
$$n_0mn-n_0m(n_0-1)\le 2 mn-(n_0-1).$$
If we move all terms to the right hand side, we get
$$n_0(n_0-1)m-((n_0-2)m+1)n+(n_0-1)\ge 0.$$
Finally, the above inequality could be rewritten as
$$
((n_0-2)m+1)\left(\frac{n_0(n_0-1)}{n_0-2}-n\right)
\ge \frac{2n_0-2}{n_0-2}$$
That is, 
\begin{equation}
\label{equation1}
n\le \frac{n_0-1}{n_0-2}\left(
n_0-\frac{2}{(n_0-2)m+1}
\right)
\end{equation}
\hfill{$\Box$}

Based on equation (\ref{equation1}), we get the necessary 
conditions for $n$ for different $n_0$ in Table \ref{necessn}.
\begin{table}[htb] 
\begin{center}
\caption{Necessary conditions for $n$ for different $n_0$}
\label{necessn}
 \begin{tabular}{ |c|c|c|}   \hline
$n_0$ &$m$ & $n$\\ \hline
3 & $[1,2]$ & 4\\ \hline
3 & $[3,\infty]$ & 5\\ \hline
$[4,\infty]$ & $[1,\infty]$ & $n_0+1$\\ \hline  
\end{tabular}
\end{center}
\end{table}
The values in Table \ref{necessn} show that for degree one and two encoding
symbols based array BP-XOR codes, if we want to recover
the information symbols from more than three columns (i.e., $n_0\ge 3$)
of encoding symbols, then we could only have one column redundancy
for $n\ge 4$.

Combining Theorem \ref{necessaryarraybpxor}
and values in Table \ref{necessn}, we get the following results
for edge-colored graphs.
\begin{theorem}
\label{edgetheorem}
For a color set $C$ with $|C|\ge 5$, if we want to design an edge-colored
graph  $G(V,E,C,f)$ (or a network with more than $|C|$ kinds of homogeneous 
devices) with minimum cost, then the edge-colored graph is robust 
against at most one color failures (or one brand of homogeneous devices failures).
\end{theorem}

{\em Proof.}
Based on the results in Theorems \ref{graphtobpxorthm}, 
\ref{necessaryarraybpxor} and values in Table \ref{necessn}, 
we can have the following conclusion:
Given integers $n_0$, $m$ and $n$,  an edge-colored graph
$G(V,E,C,f)$ with  $|C|=n$, $|V|=n_0m+1$, and $|E|=nm-(n-n_0)$,
$G(V,E,C,f)$  is $(n-n_0)$-color connected only $n=n_0+1$.
Thus the theorem follows.
\hfill$\Box$

\section{Conclusion}
Based on the BP (Belief Propagation) decoding process 
and the edge-colored graph model \cite{Wang:2011:EGA:1975023.1975459}, 
we introduced flat BP-XOR codes and array BP-XOR codes. We have established
the equivalence between edge-colored graphs and degree one and two based
array BP-XOR codes. In particular, we used results in array BP-XOR codes to 
get new results in edge-colored graphs.  For array BP-XOR codes with
higher degree encoding symbols, we do not have general results yet.
It would be interesting to have a compelete characterization of the existence
and bounds for array BP-XOR codes with
higher degree encoding symbols. These characterizations may be used 
to design more efficient LT codes or digital fountain techniques.
We have implemented an online 
software package for users to generate array BP-XOR codes
with their own specification and to verify the validity of 
their array BP-XOR codes (see \cite{bpxorurl}).

\section*{Acknowledgements}
I would like to thank Duan Qi for some discussion on Hamming code
and Theorem \ref{2erasurefaults} and thank Prof. Doug Stinson and Yvo Desmedt,
for some discussions on edge-colored graphs,
Hamiltonian circuit,  and factorization of complete graphs.

\bibliographystyle{plain}

\begin{thebibliography}{10}

\bibitem{Acedanski05howgood}
S.~Acedanski, S.~Deb, M.~MÈdard, and R.~Koetter.
\newblock How good is random linear coding based distributed networked storage.
\newblock In {\em NetCod}, 2005.

\bibitem{AndersonTopology}
B.A. Anderson.
\newblock Symmetrygroups of some perfect 1-factorizations of complete graphs.
\newblock {\em Discrete Mathematics}, 18(3):227--234, 1977.

\bibitem{DBLP:journals/tc/BlaumBBM95}
M.~Blaum, J.~Brady, J.~Bruck, and J.~Menon.
\newblock {EVENODD}: An efficient scheme for tolerating double disk failures in
  raid architectures.
\newblock {\em IEEE Trans. Computers}, 44(2):192--202, 1995.

\bibitem{Blaum96mdsarray}
M.~Blaum, J.~Bruck, and E.~Vardy.
\newblock {MDS} array codes with independent parity symbols.
\newblock {\em IEEE Trans. on Information Theory}, 42:529--542, 1996.

\bibitem{DBLP:journals/tit/BlaumR93}
M.~Blaum and R.~M. Roth.
\newblock New array codes for multiple phased burst correction.
\newblock {\em IEEE Trans. on Information Theory}, 39(1):66--77, 1993.

\bibitem{Blaum99onlowest-density}
M.~Blaum and R.~M. Roth.
\newblock On lowest-density {MDS} codes.
\newblock {\em IEEE Trans. on Information Theory}, 45:46--59, 1999.

\bibitem{CaoYuYangLouHouLTcodes}
N.~Cao, S.~Yu, Z.~Yang, W.~Lou, and T.~Hou.
\newblock Lt codes-based secure and reliable cloud storage service.
\newblock In {\em Proceedings of INFOCOM}, 2012.

\bibitem{Cassuto:2009:CLD:1669446.1669469}
Yuval Cassuto and Jehoshua Bruck.
\newblock Cyclic lowest density mds array codes.
\newblock {\em IEEE Trans. Inf. Theor.}, 55(4):1721--1729, April 2009.

\bibitem{DBLP:conf/fast/CorbettEGGKLS04}
P.~Corbett, R.~English, A.~Goel, T.~Grcanac, S.~Kleiman, J.~Leong, and
  S.~Sankar.
\newblock Row-diagonal parity for double disk failure correction.
\newblock In {\em FAST}, pages 1--14, 2004.

\bibitem{Dimakis:2010:NCD:1861840.1861868}
A.~Dimakis, P.~Godfrey, Y.~Wu, M.. Wainwright, and K.~Ramchandran.
\newblock Network coding for distributed storage systems.
\newblock {\em IEEE Trans. Inf. Theor.}, 56(9):4539--4551, 2010.

\bibitem{DBLP:journals/corr/abs-1004-4438}
A.~Dimakis, K.~Ramchandran, Y.~Wu, and C.~Suh.
\newblock A survey on network codes for distributed storage.
\newblock {\em CoRR}, abs/1004.4438, 2010.

\bibitem{Dimakis06decentralizederasure}
R.~Dimakis, V.~Prabhakaran, and K.~Ramch.
\newblock Decentralized erasure codes for distributed networked storage.
\newblock {\em IEEE Trans. Inf. Theor.}, 52:2809--2816, 2006.

\bibitem{Elias}
P.~Elias.
\newblock Error--free coding.
\newblock Technical Report 285, Massachusetts Institute of Technology (Boston),
  1954.

\bibitem{ZaiZinSem83}
N.~V.~Semakov G.~V.~Zaitsev, V.~A.~Zinov'ev.
\newblock Minimum-check-density codes for correcting bytes of errors, erasures,
  or defects.
\newblock {\em Problems Inform. Transmission}, 19(3):197--204, 1983.

\bibitem{gallagerldpc}
R.~G. Gallager.
\newblock {\em Low density Parity Check Codes}.
\newblock MIT Press, 1963.

\bibitem{Greenan:2010:FXE:1913798.1914423}
K.~Greenan, X.~Li, and J.~Wylie.
\newblock Flat xor-based erasure codes in storage systems: Constructions,
  efficient recovery, and tradeoffs.
\newblock In {\em Proc. MSST}, pages 1--14. IEEE Computer Society, 2010.

\bibitem{Huang05anefficient}
C.~Huang and L.~Xu.
\newblock {STAR:} an efficient coding scheme for correcting triple storage node
  failures.
\newblock In {\em FAST}, pages 197--210, 2005.

\bibitem{DBLP:journals/gc/Kobayashi89}
M.~Kobayashi.
\newblock On perfect one-factorization of the complete graph ${K}_{2p}$.
\newblock {\em Graphs and Combinatorics}, 5(1):351--353, 1989.

\bibitem{Krawczyk:1993:DFS:164051.164075}
Hugo Krawczyk.
\newblock Distributed fingerprints and secure information dispersal.
\newblock In {\em Proc. PODC}, pages 207--218. ACM Press, 1993.

\bibitem{DBLP:conf/focs/Luby02}
M.~Luby.
\newblock {LT} codes.
\newblock In {\em Proc. FOCS}, pages 271--280, 2002.

\bibitem{Luby:1997:PLC:258533.258573}
M.~Luby, M.~Mitzenmacher, M.~Shokrollahi, D.~Spielman, and V.~Stemann.
\newblock Practical loss-resilient codes.
\newblock In {\em Proc. 29th ACM STOC}, pages 150--159. ACM, 1997.

\bibitem{pearl}
J.~Pearl.
\newblock {\em {Probabilistic Reasoning in Intelligent Systems: Networks of
  Plausible Inference}}.
\newblock Morgan Kaufmann, 1988.

\bibitem{ldpc}
J.~Plank and M.~Thomason.
\newblock A practical analysis of low-density parity-check erasure codes for
  wide-area storage applications.
\newblock In {\em DSN}, pages 115--124, 2004.

\bibitem{Rabin:1989:EDI:62044.62050}
Michael~O. Rabin.
\newblock Efficient dispersal of information for security, load balancing, and
  fault tolerance.
\newblock {\em J. ACM}, 36(2):335--348, April 1989.

\bibitem{Verhoeff:1987:UTM:41075.41082}
Tom Verhoeff.
\newblock An updated table of minimum-distance bounds for binary linear codes.
\newblock {\em IEEE Trans. Inf. Theor.}, 33(5):665--680, September 1987.

\bibitem{bpxorurl}
Yongge Wang.
\newblock {Array BP-XOR Code Generation and Verification Webpage}.
\newblock \url{http://coitweb.uncc.edu/~yonwang/bpxor/}, 2012.

\bibitem{Wang:2011:EGA:1975023.1975459}
Yongge Wang and Yvo Desmedt.
\newblock Edge-colored graphs with applications to homogeneous faults.
\newblock {\em Inf. Process. Lett.}, 111(13):634--641, July 2011.

\bibitem{Weatherspoon:2002:ECV:646334.687814}
H.~Weatherspoon and J.~Kubiatowicz.
\newblock Erasure coding vs. replication: A quantitative comparison.
\newblock In {\em Revised Papers from the First IPTPS'01}, pages 328--338.
  Springer-Verlag, 2002.

\bibitem{Xu98lowdensity}
L.~Xu, V.~Bohossian, J.~Bruck, and D.~Wagner.
\newblock Low density mds codes and factors of complete graphs.
\newblock {\em IEEE Trans. Inf. Theor.}, 45:1817--1826, 1998.

\bibitem{Xu99x-code:mds}
L.~Xu and J.~Bruck.
\newblock X-code: Mds array codes with optimal encoding.
\newblock {\em IEEE Trans. on Information Theory}, 45:272--276, 1999.

\end{thebibliography}

\end{document}